\UseRawInputEncoding

\documentclass[aps,twocolumn,nofootinbib]{revtex4}

\usepackage{graphicx}
\usepackage{amssymb, amsmath,mathtools}
\usepackage{bbm}
\usepackage{enumerate}

\graphicspath{{./}{./plots/}}

\usepackage{xcolor}

\begin{document}

\title{$SO(8)$ unification and the large-N theory of superconductor-insulator transition of two-dimensional Dirac fermions}

\author{Igor F. Herbut and Subrata Mandal}

\affiliation{ Department of Physics, Simon Fraser University, Burnaby, British Columbia, Canada V5A 1S6}

\begin{abstract}
Electrons on honeycomb or pi-flux lattices obey effective massless Dirac equation at low energies and at the neutrality point, and should suffer quantum phase transitions into various Mott insulators and superconductors at strong two-body interactions. We show that 35 out of 36 such order parameters that provide Lorentz-invariant mass-gaps to Dirac fermions can be organized into a single irreducible tensor representation of the $SO(8)$ symmetry of the two-dimensional Dirac Hamiltonian for the spin-1/2 lattice fermions. The minimal interacting Lagrangian away from the neutrality point has the $SO(8)$ symmetry reduced to $U(1) \times SU(4)$ by finite chemical potential, and it allows only two independent interaction terms. When the Lagrangian is nearly $SO(8)$-symmetric and the ground state insulating at the neutrality point, we argue it turns superconducting at the critical value of the chemical potential through a ``flop" between the tensor components. The theory is exactly solvable when the $SU(4)$ is generalized to $SU(N)$ and $N$ taken large. A lattice Hamiltonian that may exhibit this transition, parallels with the Gross-Neveu model, and applicability to related electronic systems are briefly discussed.
\end{abstract}

\maketitle

Graphene at the neutrality point is a weakly interacting gapless Dirac semimetal. Multilayer graphene structures, in contrast, often appear to be Mott  insulators. The insulating state in rhombohedral trilayer graphene, for example, with doping (and electric field) turns into a superconductor.\cite{zhou}  The latter phenomenon is reminiscent of the much-studied but still incompletely understood behavior of cuprate superconductors. Cuprates are insulating antiferromagnets at half-filling which become d-wave superconductors with high critical temperatures above critical doping with holes. Zhang \cite{zhang, demler} viewed this insulator-superconductor transition as a ``flop" of a five-dimensional vector order parameter comprised of the three N\' eel and two superconducting components, which is induced by the chemical potential which favors the superconducting directions. Unfortunately, the $SO(3) \times SO(2)$-symmetric three-dimensional Ginzburg-Landau theory is now understood  \cite{calabrese, herbutbook} not to exhibit a particularly wide crossover regime near its unstable $SO(5)$-symmetric fixed point, as was originally hoped, which significantly reduces the range of relevance of such a  unified theory. We argue below, however, that a different unification of physically disparate orders under the umbrella of an emergent internal symmetry becomes possible in two-dimensional  (2d) Dirac systems.

As a paradigmatic example we take the electrons in graphene, which at low energies, at the neutrality point, and when assumed non-interacting, are described by the eight-dimensional Dirac Hamiltonian. The number eight comes from the honeycomb lattice being bipartite (two sublattices), there being two inequivalent Dirac cones, or valleys (fermion doubling), and finally the electrons having spin-1/2. By a judicious construction of the Dirac fermion the non-trivial matrix structure of the Dirac Hamiltonian can be completely stowed into the sublattice  factor space, and then replicated four times for the two valley- and the two spin-components. When written like this the Dirac single-particle Lagrangian besides its hallmark space-time $SU(2)$ Lorentz symmetry clearly displays the  internal $SU(4)$ symmetry, the latter acting on the spin-valley index. Of course, since the particle number is conserved, there is also the exact $U(1)$ gauge symmetry and the discrete time reversal symmetry. Both the Lorentz $SU(2)$ and the internal $SU(4)$ symmetries are emergent at low energies and broken by the lattice.

What is the minimal {\it interacting} Lagrangian that would respect the emergent  $U(1)\times SU(4) \times SU(2)$? \cite{hjr}  We show first that such a Lagrangian is remarkably simple, and contains only two independent local quartic terms. Furthermore, the 36 Dirac fermion bilinears \cite{qed31, qed32, mudry} that transform as Lorentz singlets, i. e. the average of which would be the order-parameters that represent Dirac masses, can be understood as irreducible representations (irreps) $10$, $10^*$, $15$ and $1$ under the $SU(4)$ transformations.\cite{comment} The first two irreps correspond to $x-$ and $y$-components of the 10 gapped superconducting order parameters, \cite{comment1, honerkamp, royherbut} and transform as the symmetric second-rank tensor and its complex conjugate. \cite{zee} The irrep $15$ is the adjoint representation comprising of all insulating  mass-order-parameters other than the quantum anomalous Hall  (QAH) state.\cite{haldane} The latter transforms as an $SU(4)$ singlet, i. e. as the irrep $1$.

Our principle  observation is that when a certain condition between the coupling constants of the two interaction terms is met, and when the chemical potential is at the Dirac point, the minimal {\it interacting} Lagrangian exhibits further enlargement from $U(1)\times SU(4)\times SU(2)$ symmetry to $SO(8)\times SU(2)$. The $SO(8)$ arises because the two-component 2d massive Dirac equation can be transformed into a ``real" (Majorana) form, so that in the case of graphene four copies of the usual two-component complex Dirac fermions are equivalent to eight copies of two-component Majorana  fermions. The emergent $SO(8)$ symmetry simultaneously unifies {\it almost all} the 36 mass-order-parameters as:
\begin{equation}
10 + 10^* + 15 + 1 \rightarrow 35 + 1,
\end{equation}
where on the (left-hand) right-hand side we mean the irreps of ($SU(4)$) $SO(8)$. $35$ stands for the irreducible, symmetric, second-rank $SO(8)$ tensor, and gathers together all the insulating and the superconducting mass-gaps other than the QAH state. The $SO(8)$-symmetric theory has a {\it single} interaction coupling constant, and we argue that, at strong coupling, and depending on its sign, the ground state: 1) either through the canonical Gross-Neveu transition \cite{hjr} becomes the QAH insulator, which preserves $SO(8)$ and spontaneously breaks the $Z_2$ (time reversal) symmetry, or, 2) spontaneously breaks $SO(8)$ down to $SO(4) \times SO(4)$.

As the first application of the $SO(8)$ unified theory we demonstrate that the insulator-to-superconductor transition becomes induced by the chemical potential. Taking the sign of the interaction coupling in the $SO(8)$-symmetric theory so that $SO(8)\rightarrow SO(4) \times SO(4)$ transition is realized, we show that the theory becomes exactly solvable in the large-N limit, when the original $SU(4)$ is generalized to $SU(N)$. At the neutrality point and strong coupling the exact ground state may then be any of the gapped insulators (other than QAH) or the superconductors, including some of their linear combinations. To study the competition between the insulating and superconducting ground states we detune the above condition between the two interaction couplings in the $SU(N)$ theory to explicitly favor the insulating solution at the Dirac point. Finite chemical potential is then shown to benefit the superconducting solution, and to eventually cause a first-order transition between the two competing classes of states at its critical value.

{\bf Dirac Lagrangian} - The low-energy Dirac Lagrangian of the tight-binding nearest-neighbor-hopping Hamiltonian on the bipartite honeycomb lattice and at half filling can be written as \cite{herbutprl, herbutfisherbook}
\begin{equation}
L_0=\psi^\dagger( 1_4 \otimes (1_2 \partial_\tau -i \sigma_1 \partial_1 -i \sigma_2 \partial_2 ))   \psi,
\end{equation}
where $\sigma_i$ are the standard Pauli matrices, $1_N$ is a $N$-dimensional unit matrix, and the eight-component Grassmann field $\psi^T = ( \psi_+ ^T (x), \psi_- ^T (x) )$, with $\psi_\sigma (x)=\int d^Dq e^{iqx}\psi_\sigma(q)$  given by $\psi_\sigma ^\dagger(q)=\left[u_\sigma ^\dagger(K+q),v_\sigma ^\dagger(K+q), i v_\sigma ^\dagger(-K+q), -i u_\sigma ^\dagger(-K+q)\right]$. $u_\sigma$ and $v_\sigma$ as electron variables on the triangular sublattices of the honeycomb lattice, and $\sigma=\pm$ is the third projection of spin-1/2. The $D=2+1$-dimensional energy-momentum vector $q=(\omega,\vec{q})$ collects together the Matsubara frequency $\omega$ and the wavevector $\vec{q}$, $K = (0, \vec{K})$, with $\pm \vec{K}$ as the inequivalent Dirac points, $|\vec{q}| <\Lambda \ll |\vec{K}|$, $\Lambda$ is the momentum cutoff, and $\tau$ represents the imaginary time. The reference frame is chosen so that $q_x=\vec{q}\cdot \vec{K}/|\vec{K}|$, $q_y=(\vec{K}\times\vec{q})\times \vec{K}/|\vec{K}|^2$, and the Fermi velocity is set to unity.

Let us list some global symmetries of $L_0$:

\noindent
1) gauge $U(1)$,
\begin{equation}
\psi \rightarrow e^{i\phi} \psi, \psi^\dagger \rightarrow e^{-i\phi} \psi^\dagger,
\end{equation}

\noindent
2) antiunitary time-reversal,
\begin{equation}
\psi \rightarrow (\sigma_2\otimes \sigma_2 \otimes \sigma_2 ) \psi^{*}, \psi^\dagger \rightarrow \psi^T (\sigma_2\otimes \sigma_2 \otimes \sigma_2),
\end{equation}
\vspace{1mm}

\noindent
3) internal $U \in SU(4)$,
\begin{equation}
\psi \rightarrow (U\otimes 1_2) \psi, \psi^\dagger \rightarrow \psi^\dagger (U^\dagger \otimes 1_2),
\end{equation}

\noindent
4) Lorentz $U\in SU(2)$,
\begin{equation}
\psi \rightarrow (1_4 \otimes U ) \psi, \bar{\psi} \rightarrow \bar{\psi} (1_4 \otimes U^\dagger),
\end{equation}

\noindent
accompanied by the corresponding rotation of the space-time vector $(\tau, \vec{x})$. Here, $\bar{\psi} = \psi^\dagger (1_4\otimes \sigma_3)$.

{\bf Interaction Lagrangian} - Next, we exhibit all local interaction terms quartic in fermion fields that would respect the above symmetries, with the $SU(4)$ generalized to $SU(N)$. There are four such terms:
\begin{equation}
I_1 = (\bar{\psi} (1_4 \otimes 1_2) \psi)^2, I_2 = (\bar{\psi} (G_a \otimes 1_2) \psi)^2,
\end{equation}
\begin{equation}
I_3 =  (\bar{\psi} (1_4 \otimes \sigma_i) \psi)^2, I_4= (\bar{\psi} (G_a \otimes \sigma_i) \psi)^2.
\end{equation}
Here $G_a$, $a=1,...N^2 -1$ are the hermitian generators of $SU(N)$, which satisfy $Tr (G_a G_b) = N \delta_{ab}$. The usual summation convention is assumed.

For any $N$, only two, and any two, of these four terms are linearly independent. \cite{suppl} We may chose these two to be $I_1$ and $I_2$, so that the general $U(1)\times SU(N)\times SU(2)$-invariant local interaction term in the Lagrangian becomes
\begin{equation}
L_1 = g_1 I_1 + g_2 I_2.
\end{equation}
When $g_2=0$, the Lagrangian $L= L_0 + L_1$ is nothing but the canonical Gross-Neveu model in $D=2+1$ with $N$ fermion flavors. For $N\rightarrow \infty $ and $g_1 <g_{1c}=-\pi/(4N\Lambda) <0$ one finds $\langle \bar{\psi} \psi \rangle \neq 0 $,\cite{klimenko, rosenstein, hands, kogut}  i. e. QAH state. For $g_{1c}<g_1<0$, $\langle \bar{\psi} \psi \rangle = 0$.

In \cite{suppl} we derive the following  identity:
\begin{equation}
- (N+1) I_1 = (\psi^\dagger (S_b \otimes \sigma_2) \psi^{*} ) (\psi^T (S_b  \otimes \sigma_2) \psi ) + I_2.
\end{equation}
The index $b=1,...N(N+1)/2$, and $S_b$ are symmetric, N-dimensional real matrices, with $Tr (S_a S_b) = N \delta_{ab}$. The interaction term $L_1$ can therefore also be written as
\begin{widetext}
\begin{equation}
L_1 = -\frac{g_1}{N+1} (\psi^\dagger (S_b \otimes \sigma_2) \psi^{*} ) (\psi^T (S_b  \otimes \sigma_2) \psi ) + (g_2 -\frac{g_1}{N+1})(\psi^\dagger  (G_a \otimes \sigma_3) \psi)^2.
\end{equation}
\end{widetext}
Let us now assume $g_1>0$,  $\tilde{g}_1= g_1 - (N+1) g_2  >0$, so that QAH state is suppressed, and $\langle \bar{\psi} \psi \rangle =0$. Using the Hubbard-Stratonovich (HS) transformation \cite{negele} the Lagrangian $L=L_0+L_1$ can now be expressed as
\begin{widetext}
\begin{equation}
L= L_0+ \frac{N+1}{4} (\frac{\Delta_b ^* \Delta_b}{g_1} + \frac{m_a m_a}{ \tilde{g}_1 } ) - m_a (\psi^\dagger  (G_a \otimes \sigma_3) \psi) + \frac{\Delta_b}{2}  (\psi^\dagger (S_b \otimes \sigma_2) \psi^{*} ) + \frac{\Delta_b ^* }{2}  (\psi^T (S_b  \otimes \sigma_2) \psi ).
\end{equation}
\end{widetext}
The averages of the HS fields satisfy $(N+1) \langle \Delta_b \rangle = 2 g_1 \langle \psi^T (S_b  \otimes \sigma_2) \psi \rangle $, and $(N+1) \langle m_a \rangle =  2 \tilde{g}_1 \langle \psi^\dagger  (G_a \otimes \sigma_3) \psi \rangle $, and transform as the irreps $N(N+1)/2$ (symmetric tensor) and $N^2 -1$ (adjoint) of the $SU(N)$, respectively. Both are singlets under the Lorentz $SU(2)$. When $g_2 =0$  Eq. (12) provides an alternative representation of the Gross-Neveu model.

{\bf Majorana representation} - When $g_2=0$ the symmetry of the Lagrangian $L$ is in fact $SO(2N)$. To see this, rotate $\psi \rightarrow \chi = 1_N \otimes e^{ i (\pi/4) \sigma_1} \psi$, which makes the Dirac Hamiltonian for $\chi$ fully imaginary, and transforms the fermion  bilinears as
\begin{equation}
\psi^\dagger  (G_a \otimes \sigma_3) \psi \rightarrow  \chi ^\dagger  (G_a \otimes \sigma_2) \chi,
\end{equation}
\begin{equation}
\psi^T (S_b  \otimes \sigma_2) \psi \rightarrow \chi^T (S_b  \otimes \sigma_2) \chi.
\end{equation}
The second bilinear's form does not change, since the antisymmetric $SU(2)$ tensor $\sigma_2$ transforms as a singlet.

We decompose the new Dirac fermion as $\chi = \chi_1 - i \chi_2$,  and $\chi^\dagger = \chi_1 ^T  + i \chi_2 ^T $, where $\chi_{1,2}$ are ``real", or  Majorana fermions. In terms of the $4N$-component Majorana fermion $\phi^T = (\chi_1 ^T, \chi_2 ^T)$ the Lagrangian for $g_2=0 $ now simplifies into
\begin{widetext}
\begin{eqnarray}
L= \phi^T (1_{2N} \otimes (1_2 \partial_\tau -i \sigma_1 \partial_1 -i \sigma_3 \partial_2) )  \phi +
\frac{N+1}{8N g_1} Tr S^2 + \phi^T (S\otimes \sigma_2) \phi,
\end{eqnarray}
\end{widetext}
where the matrix
\begin{equation}
S= \Delta' _{b} \sigma_3 \otimes S_b + \Delta'' _{b} \sigma_1 \otimes S_b + m_c 1_2 \otimes G_c ^{S} + m_d \sigma_2 \otimes G_d ^{A},
\end{equation}
is the general $2N$-dimensional symmetric, real, traceless matrix, and $\Delta_b = \Delta' _{b} - i \Delta'' _{b} $. $G_c ^{S}$ are the symmetric, and $G_d ^{A}$ are the antisymmetric generators of $SU(N)$, so the indices $c=1,...(N-1)(N+2)/2$, and $d= 1,...N(N-1)/2 $. The matrices $G_c ^{S}$  can be chosen  the same as the matrices $S_b$, without the unit matrix. $G_d ^A $ are then the adjoint irrep of $SO(N)\subset SU(N)$.

The Lagrangian $L$ in Eq. (15) is now manifestly invariant under the transformation $\phi \rightarrow (O\otimes 1_2) \phi$, $S\rightarrow O S O^T$, where $O \in SO(2N)$. The $(N+1) (2 N-1)$ HS fields $(\Delta' _{b}, \Delta'' _{b}, m_a)$ transform as the {\it symmetric, traceless, second-rank tensor} under $SO(2N)$.

{\bf Large-N} - In the limit $N\rightarrow \infty$ in Eq. (12) the theory becomes exactly solvable by the saddle-point method. Hereafter we take $N=2^n$ and $n$ integer. At the $SO(2N)$-symmetric point, for $g_2=0$ the minimum  is the matrix $S$ such that $S^2 = M^2 1_{2N}$, and
 \begin{equation}
 M= \Lambda- \frac{\pi}{g_1}
 \end{equation}
for $g_1 \Lambda/\pi >1$, and otherwise zero.\cite{suppl} When $M\neq 0 $ the $SO(2N)$ symmetry becomes broken to $SO(N)\times SO(N)$. If $[S, \sigma_2 \times 1_N] =0$ the ground state is an insulator, otherwise it is a superconductor. One can also show that $S$ contains at most $N+1$  different and mutually anticommuting terms in the expansion in Eq. (16). \cite{suppl, herbut2007, herbut2012} The number of Goldstone excitations is $N^2$.

When the chemical potential $\mu>0$, the Lagrangian becomes deformed as $L\rightarrow L+ L_\mu$, and the term
\begin{equation}
L_\mu = \mu \psi^\dagger \psi= \mu \phi^T (\sigma_2 \otimes 1_N \otimes 1_2) \phi
\end{equation}
reduces $SO(2N)$ to $U(1)\times SU(N)$, and the Lorentz $SU(2)$ to $SO(2)$. We therefore restore $g_2\neq 0$ to consider the  case with the lower internal symmetry, while for simplicity still retaining the Lorentz $SU(2)$ of $L_1$. By selecting $g_2 <0$ the symmetry at $\mu=0$ is broken in favor of the insulating solution. The insulating minimum satisfies $G ^2 = m^2 1_N$, $G= m_a G_a$, with $ m= \sqrt{m_a m_a} $ determined by \cite{suppl}
\begin{equation}
\frac{\pi}{\tilde{g}_1 } = \Lambda - max (\mu, m).
\end{equation}
The solution for $m$ reduces to Eq. (17) if  $\tilde{g}_1 (\Lambda-\mu) / \pi >1$, so that $\mu < m$, but is otherwise zero. The insulating solution, obtained at $\mu=0$, would this way be suppressed completely beyond a certain value of the chemical potential.\cite{comment2}   The gauge $U(1)$ is preserved, but the internal $SU(N)$ is  spontaneously broken as $SU(N) \rightarrow U(1)\times SU(N/2)\times SU(N/2)$, leading to $N^2/2$ Goldstone excitations. For $N=4$, due to local isomorphisms  $SU(4)\simeq SO(6)$ and $SU(2) \times SU(2) \simeq SO(4)$, this can also be understood as $SO(6) \rightarrow SO(2) \times SO(4)$.

For the superconducting solution one similarly finds the matrix $S_\Delta ^2 = \Delta^2 1_N$, but with $\Delta$ determined by \cite{suppl}
\begin{equation}
\frac{\pi}{g_1} =  \Lambda - \sqrt{\mu^2 + \Delta^2} + \mu Log [ \frac{\mu}{\Delta} + \sqrt{ 1+ (\frac{\mu}{\Delta})^2} ],
\end{equation}
with the last term recognizable as the familiar Cooper log. The right-hand side is uniformly increased by a finite chemical potential, and the superconducting solution is consequently {\it enhanced} relative to its value at $\mu=0$. One can show that the subgroup of the $SU(N)$ that leaves the superconducting ground state invariant is $SO(N)$. Since the particle-number $U(1)$ is also broken, the number of Goldstone excitations is now $N(N+1)/2$.

 For the Lagrangian to be almost $SO(8)$-symmetric we assume  $0<\tilde{g}_1 - g_1\ll g_1$. The insulator-to-superconductor transition then occurs at $\mu=\mu_c$, where
 \begin{equation}
 \frac{\mu_c g_1 }{\pi} = \sqrt{\frac{N |g_2 |}{g_1}  (\frac{g_1 \Lambda}{\pi}- 1 ) }.
 \end{equation}
At $\mu=\mu_c$ the Dirac mass suffers a discontinuity: \cite{suppl}
\begin{equation}
m-\Delta = \frac{\pi N |g_2| }{2 g_1 ^2 }.
\end{equation}

{\bf Lattice model} - A lattice model where the insulator-to-superconductor transition could be observable is
\begin{equation}
H = -t \sum_{\langle i,j \rangle} c_{i}^\dagger  c_{j} + \lambda \sum_{hex.} \{ \sum_{\langle \langle i,j \in hex. \rangle \rangle } \nu_{ij } c^\dagger _{i} c_{j } + h. c. \} ^2,
\end{equation}
where $c_{i} ^\dagger = (c^\dagger _{i, +}, c^\dagger _{i, -})$ creates an electron at the site $i$ on the honeycomb lattice. Besides the first hopping term the second interaction term involves next-nearest-neighbor pairs of sites on each hexagon of the honeycomb lattice, with the phase factors $\nu_{ij} = -\nu _{ji} =\pm i$, as given by the Kane-Mele model. \cite{kane} At $ \lambda =  \lambda_{c1} < 0$ and at filling one half there should be the Dirac semimetal - QAH transition \cite{remark} described by the canonical Gross-Neveu field theory \cite{herbutprl, hjv, rosenstein}, studied by the large-N expansion \cite{vasilev, gracey1, gracey2} and the conformal bootstrap. \cite{erramilli}  It represents the singlet version of the quantum-anomalous-spin-Hall transition. \cite{hjv, liu}  At  $\lambda > 0$,  on the other hand, the relevant representation should be the Eq. (15), so that for $\lambda>\lambda_{c2} >0 $  \cite{remark} the symmetry breaking pattern is $SO(8)\rightarrow SO(4) \times SO(4)$, with the lattice terms deciding  the ground state. It is conceivable that this phase transition is discontinuous for low $N$, as typical for matrix-order-parameters.\cite{pisarski1, pisarski2, scherer} Our prediction is that such a ground state turns into a gapped superconductor with doping.

{\bf Related systems} - Any 2d $2N$-component Dirac Hamiltonian can be transformed into the Majorana form, and if the short-range interactions feature an emergent $SU(N)$ and Lorentz symmetries as well, be tuned into the $SO(2N)$-symmetric form. This includes spinless fermions hopping on honeycomb or pi-flux lattices ($N=2$), and even the quasiparticles in the d-wave superconductor ($N=4$).\cite{qed32} The interpretation of the ordered states that form the representations $1$ and $(N+1)(2N-1)$ of the $SO(2N)$ depend on the physical context. Another related example is the rhombohedral trilayer  graphene, which in the simplest approximation could be described by the Hamiltonian in Eq. (2) with the replacement (in the momentum space):  $p_1 = p \cos(\theta) \rightarrow  p^3 \cos(3 \theta)$ and $p_2 = p \sin(\theta) \rightarrow  p^3 \sin (3 \theta)$. Oddness of the single-particle Hamiltonian in space and time derivatives allows it to adopt the Majorana form, but the lack of the Lorentz symmetry necessitates three independent contact interactions. The density of states  is diverging at the neutrality point, however, and consequently the critical interaction vanishes. Details will be presented in a separate publication.

{\bf Conclusion} - In reality the internal $SU(4)$  is broken by the lattice and the Coulomb interactions to $SU(2) \times SO(2)$, with $SU(2)$ as  spin-rotations, and the $SO(2)$ related to translations. \cite{qed32, hjr}  Likewise, the Lorentz symmetry is reduced to spatial rotations alone. Such a reduced symmetry allows nine independent local interaction terms. \cite{hjr, vafek, fierz}  On the other hand, the long-range Coulomb interaction is believed to be irrelevant near Gross-Neveu-like critical points. \cite{hjv, semenoff, tang} It is conceivable that, the ultraviolet complexity of real-world Dirac systems notwithstanding, the unified $SO(8)$ theory emerges as the effective low-energy description for a not-too-unrealistic choices of two-body interactions. This theory allows an exact solution in the well defined large-N limit, in which the superconductor emerges naturally and inevitably from the insulator simply by doping through a first-order transition.

{\bf Acknowledgement} - This work has been supported by the NSERC of Canada.

\begin{widetext}
\vspace{10mm}

\section{Supplemental Material}

\subsection{Fierz rearrangements}

  In this section we derive the crucial Eq. (10) in the text. The Fierz rearrangement formula \cite{hjr} says that
\begin{equation}
(\psi_1 ^\dagger M_1 \psi_2)  (\psi_3 ^\dagger M_2 \psi_4) = - \frac{1}{N^2} Tr( M_1 X_a M_2 X_b) (\psi_1 ^\dagger X_b \psi_4) (\psi_3 ^\dagger X_a \psi_2),
\end{equation}
where $\psi_i$ is a N-component Grassmann variable,  $M_1$, and $M_2$ are two $N\times N$ matrices, and $\{ X_a \}$ provide a basis in the space of $N\times N$ matrices, which we choose to be hermitian and orthonormal: $Tr(X_a X_b) = N \delta_{ab}$. Summation over repeated indices is assumed.

Let us first apply the above formula to the simplest $SU(N)$-invariant expression $(\psi^\dagger \psi)^2$:
\begin{equation}
-N (\psi^\dagger \psi)^2 =  (\psi^\dagger \psi)^2 + (\psi^\dagger G_a \psi)^2,
\end{equation}
where $G_a$ are $N^2 -1$ generators of $SU(N)$.  Similarly, the second $SU(N)$-invariant quartic term $(\psi^\dagger G_a \psi)^2$ is
\begin{equation}
-N (\psi^\dagger G_a \psi)^2 = (N^2 - 1) (\psi^\dagger \psi)^2 - (\psi^\dagger G_a \psi)^2,
\end{equation}
which, of course, is equivalent to the first equation. There is a single, independent, finite $SU(N)$-invariant contact quartic term.

We may define a column $V^T = ( (\psi^\dagger \psi)^2, (\psi^\dagger G_a \psi)^2)$, so that the Eqs. (25) and (26) can be written together as
\begin{equation}
F_N V= -N V,
\end{equation}
where the matrix $F_N$ has the elements $F_{N, 11} = F_{N, 12}=-F_{N, 22}=1$ and $F_{N, 21} = N^2 -1$. The eigenvalues of $F_N$ are $\pm N$.

We are now ready to apply the Fierz rearrangement formula to the four $SU(N)\times SU(2)$-invariant quartic terms $I_i$, $i=1,2,3,4$, defined in the text. Defining a four-component column $W^T = ( I_1, I_2, I_3, I_4)$, the result of a straightforward but somewhat tedious computation can be expressed as
\begin{equation}
F_{N\times 2} W = -2N W,
\end{equation}
with the matrix $F_{N\times 2}$ given simply by
\begin{equation}
F_{N \times 2} = F_2 \otimes F_N.
\end{equation}
The eigenvalues of the matrix $F_{N \times 2}$ are therefore $\pm 2N$, and each eigenvalue is twice degenerate. Zero eigenvectors correspond to linearly independent combinations of the quartic terms, so there are exactly two such combinations, independently of the value of $N$. This is the first result quoted in the text.

Furthermore, obviously one can take $\bar{\psi}_i$, as defined in the text, instead of $\psi^\dagger _i$ in the Eq. (24), and then rewrite first
\begin{equation}
(\bar{\psi_1} M_1 \psi_2)  (\bar{\psi_3} M_2 \psi_4) =  -(\bar{\psi_1} M_1 \psi_2)  (\psi_4 ^T  M_2 ^T \bar{\psi}_3 ^T).
\end{equation}
Taking $M_1 = M_2 = 1_{2N}$ and Fierz-rearranging the right-hand-side then immediately yields
\begin{equation}
2 N (\bar{\psi}\psi)^2  = (\bar{\psi}  Y_a \bar{\psi}^T)( \psi^T Y_a \psi),
\end{equation}
where the matrices $Y_a$ are a hermitian basis in the space of $2N \times 2N$ matrices. Since $\psi$ is Grassmann,
\begin{equation}
 \psi^T Y_a \psi = - \psi^T Y_a ^T \psi,
\end{equation}
and only the antisymmetric matrices $Y_a$ leave a finite contribution. These, on the other hand, can be written as $S_a \otimes \sigma_2$, $A_b \otimes 1_2$, $A_b \otimes \sigma_j$, with $j\neq 2$, where $S_a$ ($A_b$) are the symmetric (antisymmetric) hermitian $N\times N$ matrices. We can therefore write
 \begin{equation}
2 N I_1 = \tilde{I}_{S} + \tilde{I}_A,
\end{equation}
where
\begin{equation}
\tilde{I}_{S} = (\bar{\psi} S_a \otimes \sigma_2 \bar{\psi}^T) (\psi^T  S_a \otimes \sigma_2 \psi),
\end{equation}
\begin{equation}
\tilde{I}_A  = (\bar{\psi} A_b \otimes i \sigma_i \sigma_2 \bar{\psi}^T) (\psi^T  A_b  \otimes (-i \sigma_2 \sigma_i) \psi).
\end{equation}
We may also observe that under Lorentz transformation $\psi \rightarrow (1_N \otimes U ) \psi$, $\bar{\psi} \rightarrow \bar{\psi} (1_N \otimes U^\dagger)$ with $U\in SU(2)$, since $U^T \sigma_2 = \sigma_2 U^\dagger$, the bilinear $ \psi^T  (S_a \otimes \sigma_2) \psi$ transforms like a Lorentz scalar, whereas the bilinear $ \psi^T  ( A_b  \otimes (-i \sigma_2 \sigma_i)) \psi$ transforms as a Lorentz vector.

All quartic terms $I_i$, $i=1,2,3,4$ can similarly be expressed in terms of the bilinears $\tilde{I}_{S}$ and $\tilde{I}_A$.\cite{fierz} It is maybe simplest to directly compute $I_3$, which gives
\begin{equation}
2 N I_3 =-3  \tilde{I}_{S} + \tilde{I}_A.
\end{equation}
Since the Eq. (28) implies that
\begin{equation}
N I_3 = -(N+2) I_1 - 2 I_2,
\end{equation}
combining Eqs. (33) and (37) gives:
\begin{equation}
2N I_2 = (N-1) \tilde{I}_S - (N+1) \tilde{I}_A.
\end{equation}
Eliminating $\tilde{I}_A$ in favor of $I_2$ in Eq. (33) then gives
\begin{equation}
(N+1)  I_1 = \tilde{I}_{S} - I_2.
\end{equation}
Recalling finally that in our case $\bar{\psi} = \psi^\dagger (1_N \otimes \sigma_3)$ gives then the Eq. (10) in the main text, with the crucial overall minus sign on the left hand side.

 \subsection{Saddle-point solution at $\mu=g_2=0$}

Integrating out the fermions yields, as usual:
\begin{equation}
L= \frac{N+1}{8 N g_1} Tr S^2 - \frac{1}{4} Tr Log [ (\omega^2 + p^2 ) 1_{2N} \otimes 1_2 + S^2 \otimes 1_2 ],
\end{equation}
where $Tr$ stands for the summation over frequency, momentum, and the diagonal elements of the matrix. Let us first assume that at the saddle-point $S^2 = M^2 1_{2N}$, and prove this assertion later. Then one finds that
\begin{equation}
\frac{N+1} { 4g_1} M= N \int_{-\infty} ^{\infty} \frac{d\omega}{2\pi} \int _{0}^{\Lambda} \frac{ d\vec{p}}{(2\pi)^2} \frac{M}{\omega^2 + p^2 + M^2},
\end{equation}
 so if $M\neq 0$,
 \begin{equation}
 \frac{\pi(N+1) }{N g_1} = \sqrt{ \Lambda^2 + M^2} - |M|.
 \end{equation}
For $N\gg 1$ and $|M| \ll \Lambda$ this reduces to the Eq. (17) in the main text.

We next show that the matrix $S^2$ indeed  at the minimum of the Lagrangian needs to be proportional to the unit matrix. Since $N=2^n$, all the basis matrices in Eq. (24) are direct products of Pauli matrices and $1_2$, so they either mutually commute or anticommute. Let us assume that $S^2$ is not proportional to unity. Then, for example, it could be
  \begin{equation}
  S= \sum_{i=1}^{K} M_i S_i + h \tilde{S},
  \end{equation}
where $\{S_i, S_j \} = 2 \delta_{ij} 1_{2N}$, $K=N+1$ is the maximal number of mutually anticommuting matrices, $ \tilde{S}^2=1_{2N}$, $\tilde{S} \neq S_i$, $i=1,...K$. To be specific, assume $ \tilde{S} $ commutes with $S_1$ and $S_2$ and anticommutes with $S_{3,4,5}$. Generalization to other cases is straightforward. Then,
\begin{equation}
S^2 = (\sum_{i=1}^{K} M_i M_i + h^2) 1_{2N} + 2h (M_1 S_1 + M_2 S_2) \tilde{S}.
\end{equation}
Since the matrices $S_1 \tilde{S}$ and $S_2 \tilde{S} $ anticommute, the N-degenerate eigenvalues of $S^2$ are
 \begin{equation}
 (\sum_{i=1}^{K} M_i M_i + h^2) \pm 2 h \sqrt{M_1 ^2 + M_2 ^2 } =  ( \sqrt{M_1 ^2 + M_2 ^2 } \pm h)^2 +  \sum_{i=3}^{K} M_i M_i.
 \end{equation}
The value of the Lagrangian for this ansatz would be
\begin{eqnarray}
L= \frac{N+1}{4 g_1} (\sum_{i=1}^{K} M_i M_i + h^2) -  \\ \nonumber
\frac{1}{4} \{ Tr Log [  ( \omega^2 + p^2 + ( \sqrt{M_1 ^2 + M_2 ^2 } +  h)^2 +  \sum_{i=3}^{K} M_i M_i )1_N \otimes 1_2 ] \\ \nonumber
+ Tr Log [  ( \omega^2 + p^2 + ( \sqrt{M_1 ^2 + M_2 ^2 } - h)^2 +  \sum_{i=3}^{K} M_i M_i )1_N \otimes 1_2 ]  \}.
\end{eqnarray}
Define the variables $M_{12,\pm} = \sqrt{M_1 ^2 + M_2 ^2 } \pm  h $, so that $2( M_1^2 + M_2 ^2 + h^2) = M_{12,+} ^2 + M_{12,-}^2$. Differentiating the Lagrangian with respect to $M_{12,\pm}$ gives that at the saddle point,
\begin{equation}
\frac{(N+1)}{4 g_1} M_{12,s} = N  \int_{\-\infty} ^{\infty} \frac{d\omega}{2\pi} \int _{0}^{\Lambda} \frac{ d\vec{p}}{(2\pi)^2} \frac{M_{12,s}}{\omega^2 + p^2 + M_{12,s} ^2 + M^2 },
\end{equation}
for $s=\pm$, and similarly for $M = (M_3^2 + M_4 ^2 + M_5 ^2)^{1/2}$,
\begin{equation}
\frac{(N+1)}{2 g_1} M = N  \sum_{s=\pm} \int_{\-\infty} ^{\infty} \frac{d\omega}{2\pi} \int _{0}^{\Lambda} \frac{ d\vec{p}}{(2\pi)^2} \frac{M}{\omega^2 + p^2 + M_{12,s} ^2  + M^2 }.
\end{equation}

 We now assume that $g_1 > \pi/\Lambda$. If all three solutions $M$ and $M_{12,\pm}$ are nontrivial, it immediately follows that  $|M_{12,+}| = |M_{12,-}|$. Likewise, if the solution is $M =0$, the Eqs. for $M_{12,+}$ and $M_{12,-}$ decouple, and minimum is at $|M_{12,+}| = |M_{12,-}|\neq 0$. This means that either $h=0$, or $M_1=M_2 =0$. In either case the matrix $S^2$ becomes proportional to unit matrix.

Another minimum is found at $|M_{12,+}| = |M_{12,-}|=0$, $M\neq0$. In this case, however, $M_1=M_2 =h=0$, and $S^2$ is again proportional to the unit matrix.

Finally, let us check if the minimum can be at  $M_{12,-}=0$, and $M_{12,+}\neq 0$ and $M\neq 0$, for example. Combining the three equations  it follows then that
\begin{equation}
\frac{(N+1)}{4 g_1}  = N  \int_{\-\infty} ^{\infty} \frac{d\omega}{2\pi} \int _{0}^{\Lambda} \frac{ d\vec{p}}{(2\pi)^2} \frac{1}{\omega^2 + p^2 + M^2}.
\end{equation}
Comparing with the equation for $M_{12,+}$ this immediately implies that $M_{12,+}=0$ as well, so we arrive at the contradiction. We have therefore proved that $S^2$ at the minimum must be proportional to the unit matrix, as claimed.

\subsection{ Saddle-point solution at $\mu>0$ and $g_2 <0$: insulator }

At the purely insulating saddle point the Lagrangian becomes
\begin{equation}
L= \frac{N+1}{4 \tilde{g}_1} m^2  - N  \int_{-\infty} ^{\infty} \frac{d\omega}{2\pi} \int _{0}^{\Lambda} \frac{ d\vec{p}}{(2\pi)^2} Log[ (\omega+i\mu)^2 + p^2 + m^2 ].
\end{equation}
Minimum for $m\neq 0$ is determined by the condition
\begin{equation}
\frac{N+1}{4 \tilde{g}_1}  =  N  \int_{-\infty} ^{\infty} \frac{d\omega}{2\pi} \int _{0}^{\Lambda} \frac{ d\vec{p}}{(2\pi)^2} \frac{1}{(\omega+i\mu)^2 + p^2 + m^2 },
\end{equation}
which after the integration over frequency gives
\begin{equation}
\frac{N+1}{4 N \tilde{g}_1} =  \int _{0}^{\Lambda} \frac{ d\vec{p}}{(2\pi)^2} \frac{\Theta ( \sqrt{ p^2 + m^2} -\mu )}{ 2 \sqrt{ p^2 + m^2} }.
\end{equation}
Taking into account the step-function in the integrand:
\begin{equation}
\frac{(N+1) \pi }{N \tilde{g}_1} =  \Theta(m-\mu) (\sqrt{ \Lambda^2 + m^2 } -m ) + \Theta ( \mu-m ) (\sqrt{\Lambda^2 + m^2} - \mu ).
\end{equation}
Assuming again $m\ll \Lambda$ and $N\gg 1$ yields the Eq. (19) in the text.

\subsection{ Saddle-point solution at $\mu>0$ and $g_2 <0$: superconductor}

At a purely superconducting saddle point the Lagrangian becomes
\begin{equation}
L= \frac{N+1}{4 g_1} |\Delta|^2  - \frac{N}{2}  \int_{-\infty} ^{\infty} \frac{d\omega}{2\pi} \int _{0}^{\Lambda} \frac{ d\vec{p}}{(2\pi)^2}
[ Log[\omega^2 + (p+\mu)^2 + |\Delta| ^2 ] + Log[\omega^2 + (p-\mu)^2 + |\Delta| ^2 ] ].
\end{equation}
Differentiating with respect to $|\Delta|$, integrating over the frequency, and assuming $|\Delta|\neq 0$ leads to
\begin{equation}
\frac{2 \pi (N+1)}{N g_1} = \int_{0}^{\Lambda} dp [ \frac{p}{\sqrt{ (p+\mu)^2 +|\Delta|^2 }  }  + \frac{p}{\sqrt{ (p-\mu)^2 +|\Delta|^2 }} ].
\end{equation}
Adding and subtracting $\mu$ in the two numerators on the right-hand-side then gives
\begin{eqnarray}
\frac{2 \pi (N+1)}{N g_1} = (\sqrt{ (\Lambda+\mu)^2 + |\Delta|^2 } - \sqrt{ \mu^2 + |\Delta|^2 } )  + (\sqrt{ (\Lambda-\mu)^2 + |\Delta|^2 } - \sqrt{ \mu^2 + |\Delta|^2 } ) +  \\ \nonumber
\mu \int_{0}^{\Lambda} dp [ \frac{1}{\sqrt{ (p-\mu)^2 +|\Delta|^2 }  }  - \frac{1}{\sqrt{ (p+ \mu)^2 +|\Delta|^2 }} ].
\end{eqnarray}
Rearranging the  the right-hand-side we get
 \begin{equation}
\frac{2 \pi (N+1)}{N g_1} = 2 (\sqrt{(\Lambda-\mu) ^2 + |\Delta|^2 } - \sqrt{ \mu^2 + |\Delta|^2 } ) +
2 \mu \int_{0}^{\mu}\frac{dp }{\sqrt{ p^2 +|\Delta|^2 }  }  .
\end{equation}
Finally, assuming again $\mu \ll \Lambda$, $|\Delta| \ll \Lambda$, and $N\gg 1$ and performing the momentum integral yields the Eq. (20) in the text.

\subsection{Superconductor-insulator transition}

 We assume $\tilde{g}_1 > g_1 > \pi/\Lambda$, so that at $\mu=0$ the insulating solution $m\neq 0$ is larger than the superconducting solution $\Delta\neq 0$, and both are much smaller that the cutoff $\Lambda$. We can then obtain the value of the Lagrangian at finite $\mu$ by integrating the Eq. (19) in the text:
 \begin{equation}
 L(m, \Delta=0) = \int_{0} ^m  m [ \frac{\pi}{\tilde{g}_1 } - \Lambda + max (\mu,m) ] dm + constant,
 \end{equation}
 which yields
 \begin{equation}
 L(m, 0) -L(0,0) = (\frac{\pi}{\tilde{g}_1} - \Lambda ) \frac{m^2}{2} +\frac{m^3}{3}.
 \end{equation}
 Inserting the $m>\mu$ solution of Eq. (19) in the main text then gives the value of the Lagrangian  at the insulating saddle-point to be
 \begin{equation}
 L(m, 0) -L(0,0) = -\frac{1}{6}  ( \Lambda- \frac{\pi}{\tilde{g}_1})^3.
 \end{equation}

 Next we do the same for the superconducting solution. Assuming that $SO(8)$ is an approximate symmetry and that therefore $\tilde{g}_1 -g_1 \ll g_1$ implies that the transition from the insulator into the superconductor will occur at $\mu = \mu_c \ll \Delta$, where $\Delta$ is the solution of Eq. (20) in the text at $\mu=0$. We can therefore expand the right-hand side of Eq. (20) in powers of $\mu$, and write
 \begin{equation}
 L(m=0, \Delta) = \int_{0} ^{\Delta} \Delta [ \frac{\pi}{g_1} -\Lambda + \Delta (1+ \frac{\mu ^2}{ 2 \Delta ^2} + O(\mu^4)) -\frac{\mu^2}{\Delta} + O(\mu^4)]
 d\Delta  + constant,
 \end{equation}
which gives
\begin{equation}
L(0,\Delta) -L(0,0) = \frac{\Delta^2}{2}(\frac{\pi}{g_1} - \Lambda )+\frac{\Delta^3}{3} - \frac{\mu^2}{2} \Delta + O(\mu^4).
\end{equation}
Inserting the superconducting solution
\begin{equation}
\Delta= \Lambda - \frac{\pi}{g_1} + \frac{\mu^2}{2(\Lambda - \frac{\pi}{g_1})} + O(\mu^4),
\end{equation}
then gives the value of the Lagrangian to be
\begin{equation}
L(0,\Delta) -L(0,0)=  -\frac{1}{6}  ( \Lambda- \frac{\pi}{g_1})^3 - \frac{\mu^2}{2} ( \Lambda- \frac{\pi}{g_1}) +O(\mu^4).
\end{equation}

The transition occurs therefore when $L(0,\Delta)= L(m,0)$, at
\begin{equation}
\mu_c ^2 = (\frac{\pi}{g_1} -\frac{\pi}{\tilde{g}_1} ) (\Lambda - \frac{\pi}{g_1}) +O( (\tilde{g}_1 - g_1)^2 ),
\end{equation}
which when slightly rewritten yields the Eq. (21) in the main text.

The discontinuity of the Dirac mass at $\mu=\mu_c$ is therefore
\begin{equation}
m-\Delta = (\Lambda - \frac{\pi}{\tilde{g}_1}) - [ (\Lambda - \frac{\pi}{g_1}) +  \frac{\mu_c ^2}{2( \Lambda- \frac{\pi}{g_1} ) } ]  +O(\mu_c ^4) ,
\end{equation}
which yields Eq. (22) in the text.

\end{widetext}


\begin{thebibliography}{99}
\bibitem{comment} Here we call an irrep by its dimension.
\bibitem{comment1} The ten superconducting order parameters are the singlet s-wave, triplet f-wave, \cite{honerkamp} and two triplet Kekule superconductors \cite{royherbut}.
\bibitem{comment2} This is what happens to the QAH solution (at $g_1 < g_{1c}$), which this way turns into the symmetric phase.
\cite{klimenko, rosenstein, hands, kogut}
\end{thebibliography}
\end{document}